# Highly Efficient and Electrically Robust Carbon Irradiated SI-GaAs Based Photoconductive THz Emitters

**Abhishek Singh, Sanjoy Pal, Harshad Surdi, S. S. Prabhu*, Vandana Nanal and R. G. Pillay**

*Tata Institute of Fundamental Research, Homi Bhabha Road, Mumbai 400005, India*

**Prabhu@tifr.res.in*

**Abstract:** We demonstrate here an efficient THz source with low electrical power consumption. We have increased the maximum THz radiation power emitted from SI-GaAs based photoconductive emitters by two orders of magnitude. By irradiating the SI-GaAs substrate with Carbon-ions up to 2 μm deep, we have created lot of defects and decreased the life time of photo-excited carriers inside the substrate. Depending on the irradiation dose we find 1 to 2 orders of magnitude decrease in total current flowing in the substrate, resulting in subsequent decrease of heat dissipation in the antenna. This has resulted in increasing maximum cut-off of the applied voltage across Photo-Conductive Emitter (PCE) electrodes to operate the device without thermal breakdown from ~35 V to > 150 V for the 25 μm electrode gaps. At optimum operating conditions, carbon irradiated ($10^{14}$ ions/cm$^2$) PCEs give THz pulses with power about 100 times higher in comparison to the usual PCEs on SI-GaAs and electrical to THz power conversion efficiency has improved by a factor of ~ 800.

---

Electromagnetic radiations having frequencies in Tera-hertz (THz) range (1THz = $10^{12}$ Hz) are not so easy to generate [1]. But due to its applications in security imaging, bio-sensing, chemical identification, material characterization etc., there is high demand of high power THz sources, particularly sources which can generate short THz pulses with broadband spectrum. Till now, photoconductive emitters (PCEs) are known to be the best sources for high power THz pulse generation. Improving the efficiency of THz pulse sources with better designs or material, is one of the major goals of ongoing research in this field. There have been several attempts to increase the THz emission from these sources by modifying the electrical and optical properties of the semiconducting substrate [2], design of electrodes [3] and patterning the active area of PCE in between the two electrodes [4].

In THz PCEs newly photo generated charge carriers (electron-hole pairs) via laser pulse of width less than 100 fs gets accelerated under already applied electric field and this sudden jump in number of free carriers and their acceleration gives sudden rise in the current. This sudden rise in current in pico-second time domain is responsible for THz pulse emission. In the case, where the semiconductor has carrier life time of less than a pico-second like LT-GaAs, current falls down to the dark level within picoseconds as electron hole pairs recombine with each other. Such materials are useful for the generation of bipolar THz pulses. In semiconductors like SI-GaAs which has carrier life time of more than 50 ps, fall in current takes relatively much longer time. Since electric field of the emitted THz pulse $E_{THz} \propto \partial J(t)/\partial t$, where *J(t)* is time dependent current density; slow varying or constant currents plays no role in THz emission. Current density can be written as $J(t) = N(t)\mu E_b$, where $N(t)$ is number of charge carriers, $\mu$ is mobility of charge carrier and $E_b$ is the applied electric field. It means that amplitude of electric fields of THz pulses emitted from PCE sources should increase linearly with number of newly generated charge carrier i.e. power of the carrier excitation pulse and with applied bias voltage across electrodes. But in practice this increase is highly sub linear and eventually saturates at higher optical powers due to screening effects and also at higher applied biases due to higher dark DC currents and heating effect caused by this current. Heat generated due to large current causes thermal breakdown of the source device, hence one cannot apply voltages beyond some limits. There have been attempts to avoid the saturation effects at higher optical power excitation using large aperture PCEs [5], [6], [7], [8] and also by attaching heat sinks with antenna electrodes to avoid the heating effect at higher bias voltages [9], [10].

In this article, we describe the effect of Carbon ion irradiation with doses of~$10^{12}$ to $10^{15}$ $cm^{-2}$ on SI-GaAs crystal. The irradiation was carried out using a 33.5 MeV beam of $^{12}C$ from the Pelletron Linac Facility, Mumbai. The beam was passed through ~10 μm thin gold foil which acted as an energy degrader and also generated the energy spread. Optimization of the gold foil thickness and incident beam energy was done using SRIM [www.srim.org] to get a nearly uniform distribution of defects up to ~2 μm depth inside the SI-GaAs crystal. The sample holder and the vacuum chamber assembly were electrically isolated from the beam-line. The incident ion dose was estimated from the observed beam spot size and the measured beam current. PCE having electrodes separated by 25 μm were fabricated using standard photolithography technique on (un-annealed) irradiated and non-irradiated substrates of SI-GaAs. These substrates were taken from the same ingot, so PCEs fabricated on the irradiated and non-irradiated substrates could be compared directly for the performance. In Fig.1 we show Current vs Voltage measurements of THz sources fabricated on usual non-irradiated SI-GaAs and carbon irradiated SI-GaAs with a dose of ~$10^{14}$ $cm^{-2}$ under dark and 125 mW (10 fs, 76 MHz, 780 nm laser pulses) illumination conditions. We observe that for 20 V of applied bias, current in the non-irradiated device goes up to 10 mA. Using simple formula for heat dissipation P = Voltage (V) × Current (I) = 0.2 J/s is the electrical power dissipated in this PCE. Assuming the volume of our PCE device ~ 5mm × 5mm × 0.35mm and SI-GaAs density as 5.3 $gm/cm^3$; its mass will be ~ 50 mg. Specific heat of GaAs is around 0.33 J/gm°C, Therefore, the heat generated around 20 V of applied bias may be large enough to affect the efficiency of the device and it may damage the device if voltage is further increased (We observed several degrees increase in the temperature of non-irradiated antenna substrate at 40 V bias even under dark conditions). Comparing current in Carbon irradiated ($10^{14}$ $cm^{-2}$) device, we find that it is less by at least two orders of magnitude under the same optical illumination condition of 125 mW. Hence in this case the heat generation will be at least two orders of magnitude smaller at 20 V applied bias.

In our case it has been observed that PCEs with electrode gap of 25 μm fabricated on usual non-irradiated SI-GaAs could not operate as efficiently as expected (This is shown in Fig.2 by black colored filled squares curve) at higher applied voltages. Electric field of the THz pulse should have been increasing linearly with applied electric field ($E_b$) according to the formula$E_{THz} \propto \partial N(t)\mu E_b/\partial t$. But the increase in electric field of the THz pulse was highly sub-linear and it was getting saturated at high bias fields. On further increasing the applied voltage eventually thermal breakdown was occurring around 35 V to 40 V bringing the THz signal to zero, destroying emitter antenna. One can easily conclude that once THz pulse electric field has stopped increasing with increase in applied voltage across electrodes, it means the thermal breakdown point is not too far and further increase in applied voltage is not recommended. Whereas in the case of PCE fabricated on carbon irradiated SI-GaAs with a dose of ~$10^{14}$ $cm^{-2}$, THz electric field was increasing linearly up to 20 V of applied voltage (Fig. 2 red filled circles curve). This improvement in device performance is attributed to significant decrease in current flowing through the device and hence subsequent decrease in the heat dissipation under same applied bias and optical illumination.

Effect of ion irradiation on carrier life time of semiconductors has been already studied [11]. It is well known that defects created by the ions, increase the scattering and charge trapping centers leading to reduction in carrier life time. To study the effect of irradiation on photo-excited carriers in SI-GaAs we also did the optical pump-probe reflectivity measurement with 10 fs, 800 nm laser pulses. The reflection signal of the probe beam changes with time as the carriers generated by the pump-beam get trapped/captured. This measurement can be used to estimate the lifetime of carriers. In Fig.3 we show 800 nm incident pump-probe reflectivity measurement curves for all the C-irradiated and non-irradiated SI-GaAs samples. From the curves it is clear that as we go on increasing the C-irradiation dose, the carrier lifetime is reduced. The curves are normalized to determine qualitatively the fall time of the reflected probe-signal. The carrier lifetime for the $10^{15}$ $cm^{-2}$ dose sample is the lowest (~0.55 ps) and for SI-GaAs (shown in the inset of Fig.3) is the highest (~70 ps). In the inset of Fig.3, for SI-GaAs, we see two exponential components, initial fast one (~3 ps) followed by a long one (~70 ps). Since the most dominant one is the long one, we quote that value. The inset also shows decay signal of the $10^{12}$ $cm^{-2}$ irradiated SI-GaAs, which has ~6 ps estimated carrier lifetime. A rough estimate of the relative mobility of the irradiated samples can be made by comparing reflection changes of the curves with respect to that of SI-GaAs, and we find that there is not much difference in the mobility of carriers in the irradiated samples in the optical frequency regime except $10^{15}$ $cm^{-2}$ dose sample. These samples are not annealed, but still have comparable mobility of carriers which is why the emission of THz is quite strong from these samples in spite of damages caused by irradiation.

In Fig.4 we show current-voltage (I-V) characteristics of all five PCE sources, namely one non-irradiated and four irradiated with doses $10^{12}$, $10^{13}$, $10^{14}$ and $10^{15}$$cm^{-2}$. It is observed that there is continuous decrease in total current flowing under illumination as we increase the irradiation dose. This suggests that heat generation will be less in the sources as we keep on increasing the irradiation dose, hence thermal breakdown voltage of PCEs will be higher at higher irradiation doses. But there is one more parameter: electron mobility ($\mu$) which plays key role in THz emission. Electric field of the emitted THz pulse is proportional to the electron mobility (hole mobility in GaAs is much smaller than electron mobility hence only electrons are major contributors in THz emission) and mobility is expected to decrease as we increase the irradiation dose. The generated defects at high enough dose will reduce the mobility of carriers by increasing scattering. Therefore we cannot keep on improving the PCE source efficiency by increasing the irradiation dose indefinitely. After some point, efficiency of source will reduce even if we can apply higher voltages without thermal breakdown. This is evident from the fact that for highest irradiation dose ($10^{15}$ $cm^{-2}$), the THz emission is lower than that from $10^{14}$ $cm^{-2}$ dose. Thus there is an optimal irradiation dose which gives maximum THz emission. The normalized (w.r.t. SI-GaAs) time resolved reflectivity curves which are shown in Fig.3, the peak signal amplitude of $10^{15}$ $cm^{-2}$ irradiated sample is significantly lower compared to others. Since this signal is proportional to the mobility of the carriers, in the optical excitation wavelength region the carrier mobility of samples up to $10^{14}$ $cm^{-2}$ dose do not show significant change, whereas sample with the highest dose ($10^{15}$ $cm^{-2}$) has shown significant decrease. To get the precise estimate of the mobility in the THz frequency regime, IR-pump-THz probe experiment is needed [11].

To study the THz power emission efficiency, PCE sources were gated with 800 nm, 10 fs, 76 MHz laser pulses and the incident optical beam was focused using an off-axis parabolic mirror of 2.0 inch focal length giving a focused beam spot of ~50 μm. The emitted THz radiation was detected using standard THz-Time domain Spectroscopy (THZ-TDS) set up with ZnTe <110> crystal for electro-optic detection. Source was electronically chopped by giving it a square wave bias voltage. Using this we recorded the time domain THz signal emitted from the antenna which allows us to study the amplitude of the Electric field as well as the frequency distribution of the emitted THz radiation. Source was tested without any THz collimating silicon lens on the other side of source to avoid the THz signal variations coming from its misalignment or dispersion effects. In Fig.5 we show variation of the peak of emitted THz field amplitude with applied bias for all five PCE sources. Non-irradiated sample sources get saturated around 20 V applied bias and maximum THz signal (which is proportional to the electric field at the pulse peak) is recorded. Thermal breakdown of non-irradiated source occurred at 35-40 V of applied voltage. Carbon irradiated source with dose of $10^{12}$ $cm^{-2}$ showed relatively linear increase in THz signal with applied bias. We were able to get maximum THz signal 5 times more than that from the non-irradiated source but at 100 V applied bias without thermal breakdown from this irradiated source. Performance of sources with irradiation doses of $10^{13}$ $cm^{-2}$ and $10^{14}$ $cm^{-2}$ was even better. Maximum THz signal we could get from $10^{13}$ $cm^{-2}$ dose source was 8 times more than that from the non-irradiated source at 110 V applied bias. Thermal breakdown of this source occurred at 110 V applied bias. With a source of $10^{14}$ $cm^{-2}$ irradiation dose, we were able to get maximum THz signal ~10 times (than that from the non-irradiated source, but) at 150 V applied bias without damaging the source. Thus in optimum operating condition (i.e. 20-30 V for non-irradiated source and 150 V for source of $10^{14}$ $cm^{-2}$ irradiation dose) THz power emitted from the irradiated source was ~100 times larger than that of usual non-irradiated SI-GaAs source. Performance of the source with irradiation dose of $10^{15}$ $cm^{-2}$ showed deteriorated mobility and maximum THz signal we could get was only larger by ~8 times (than that from the non-irradiated source, but) at 150 V applied bias. It is expected that we can apply even higher bias voltages on this source, but was not done, since the improvement in the THz signal was not expected to be any better.

To make a rough estimate of improvement in electrical to THz power conversion efficiency of the C-irradiated sources, we can compare the THz power emission at say 20 V for C-irradiated ($10^{14}$ $cm^{-2}$ dose sample) with non-irradiated sample. From Fig.2, the THz amplitude emission is almost twice (Power will be 4 times) from the irradiated sample compared to non-irradiated one. The current at this voltage is (from Fig.1) about 200 times more at the same illumination for non-irradiated sample compared to irradiated sample. Thus for the same voltage, current is 200 times less and emitted THz power is 4 times more for the irradiated sample compared to the non-irradiated sample. Thus electrical to THz power conversion efficiency of C-irradiated source is almost ~800 times better than non-irradiated source at the same applied voltage.

The Fast Fourier Transforms (FFTs) of the temporal waveforms from all the sources did not show any significant difference. Comparison with SI-GaAs sources shows that C-irradiated sources give much better performance and this is achieved by using SI-GaAs material which is easily available in open market. Compared to expensive and difficult to procure material like LT-GaAs, irradiated substrate offers a much cheaper route to better THz emitter substrate material. We are currently investigating the above antennas to be used as THz detectors, and the one with highest irradiation dose looks very promising. However, this will be reported in another article separately.

In summary, we have decreased the total current flowing through the PCE sources by irradiating the SI-GaAs, the base substrate used for making PCE device with carbon ion irradiation. This has resulted in less heat generation in the device allowing us to apply higher bias voltages across the PCE electrodes without thermal breakdown of the device. This has enabled us to get ~10 times higher THz amplitude signal (~100 times more THz power) at the same optical excitation power. Since emission of THz from PCE depends only on rise and fall of the current and not on the total average current flowing per excitation incident, we are successful in reducing the total average current by making the current fall quicker in time. This has resulted in the demonstration of better efficiency THz sources which can match LT-GaAs.

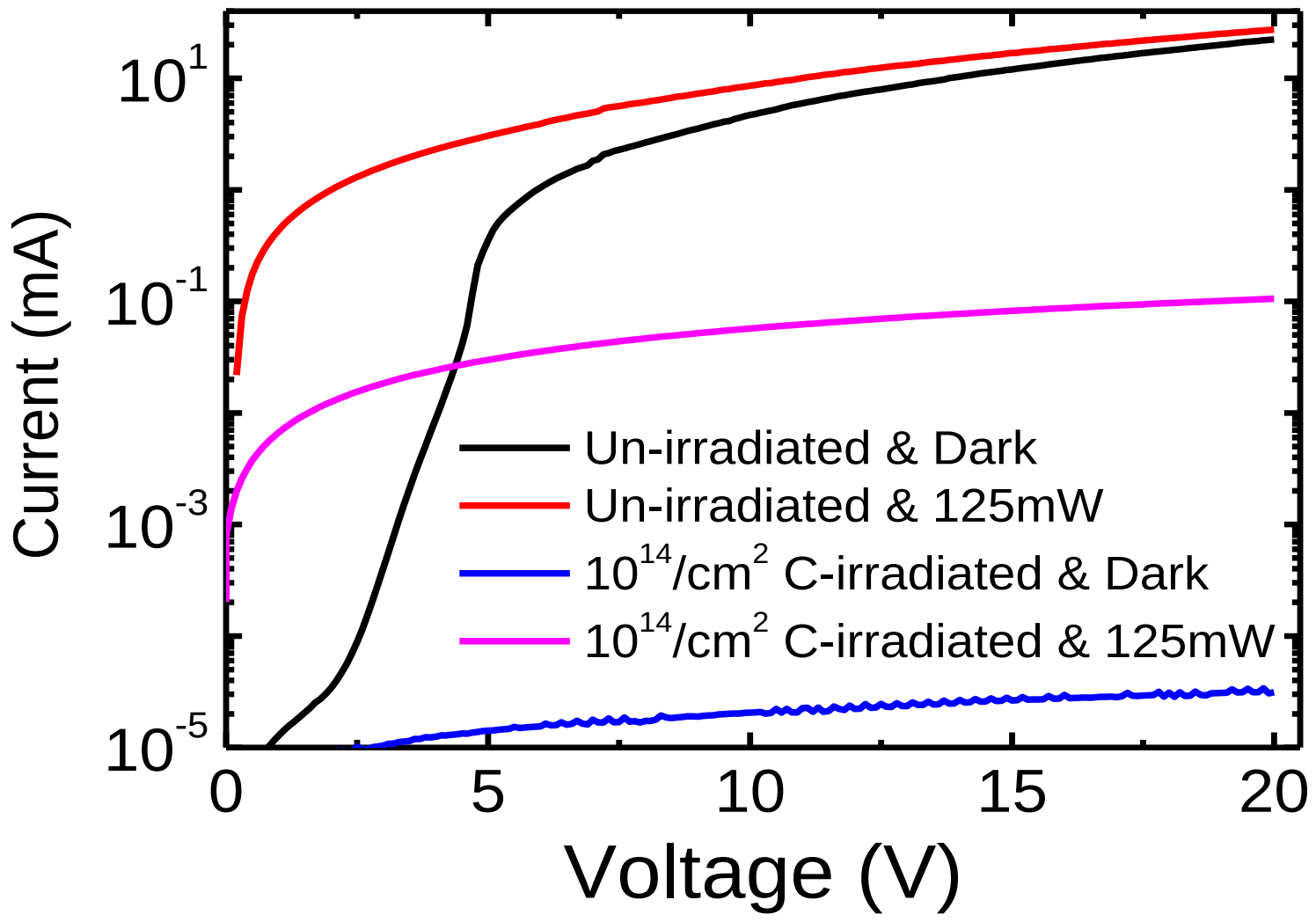

Figure1.Current-Voltage characteristics of irradiated (~$10^{14}$ cm$^{-2}$) and non-irradiated PCEs fabricated on SI-GaAs substrate under Dark and 125 mW IR-pulsed laser illumination conditions.

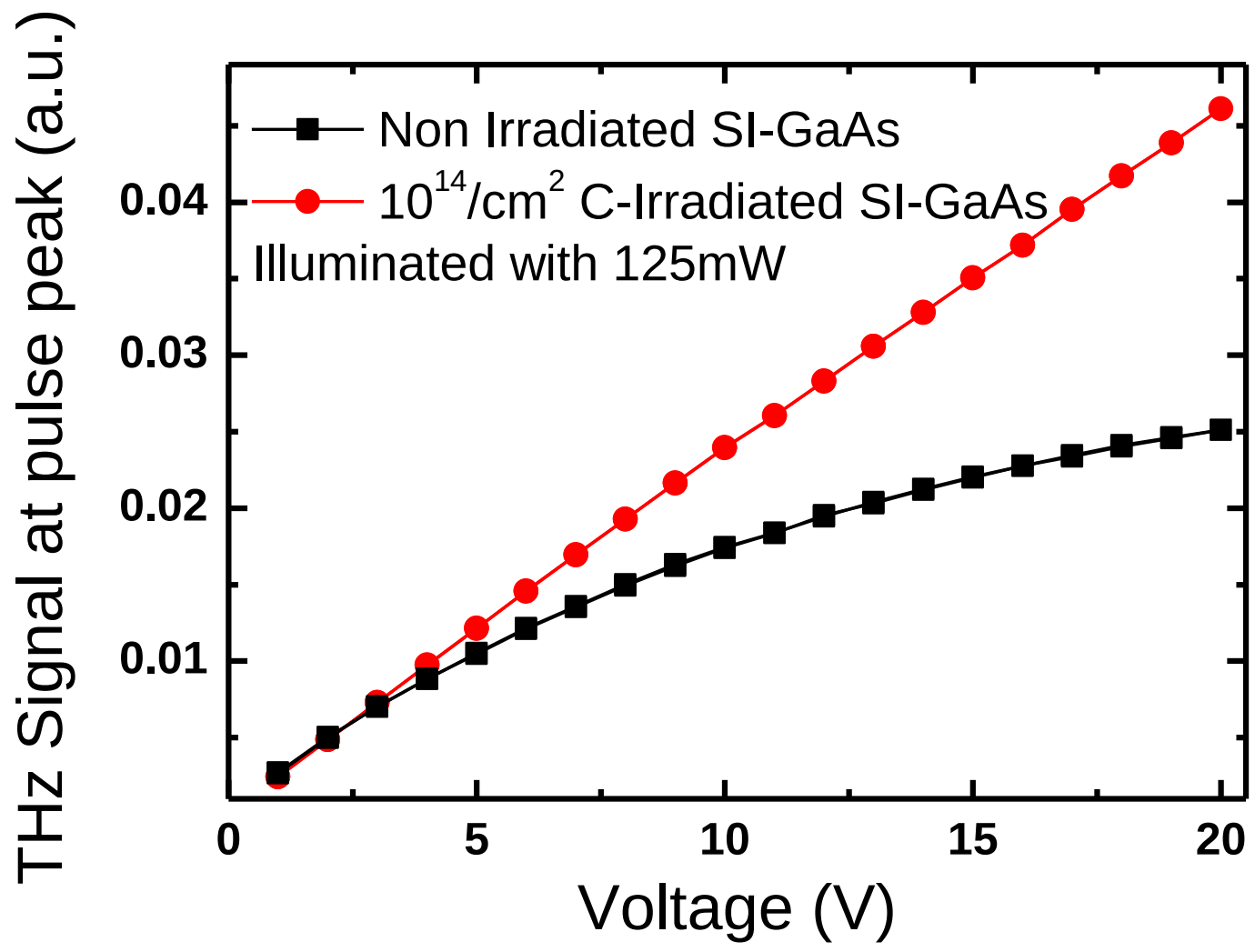

Figure2. THz Amplitude vs applied bias form usual and $10^{14}$ cm$^{-2}$ Carbon Irradiated SI-GaAs sources under 125mW illumination. The non-irradiated source shows sub-linear increase of THz amplitude (black squares) with applied bias and irradiated source shows linear increase (red circles).

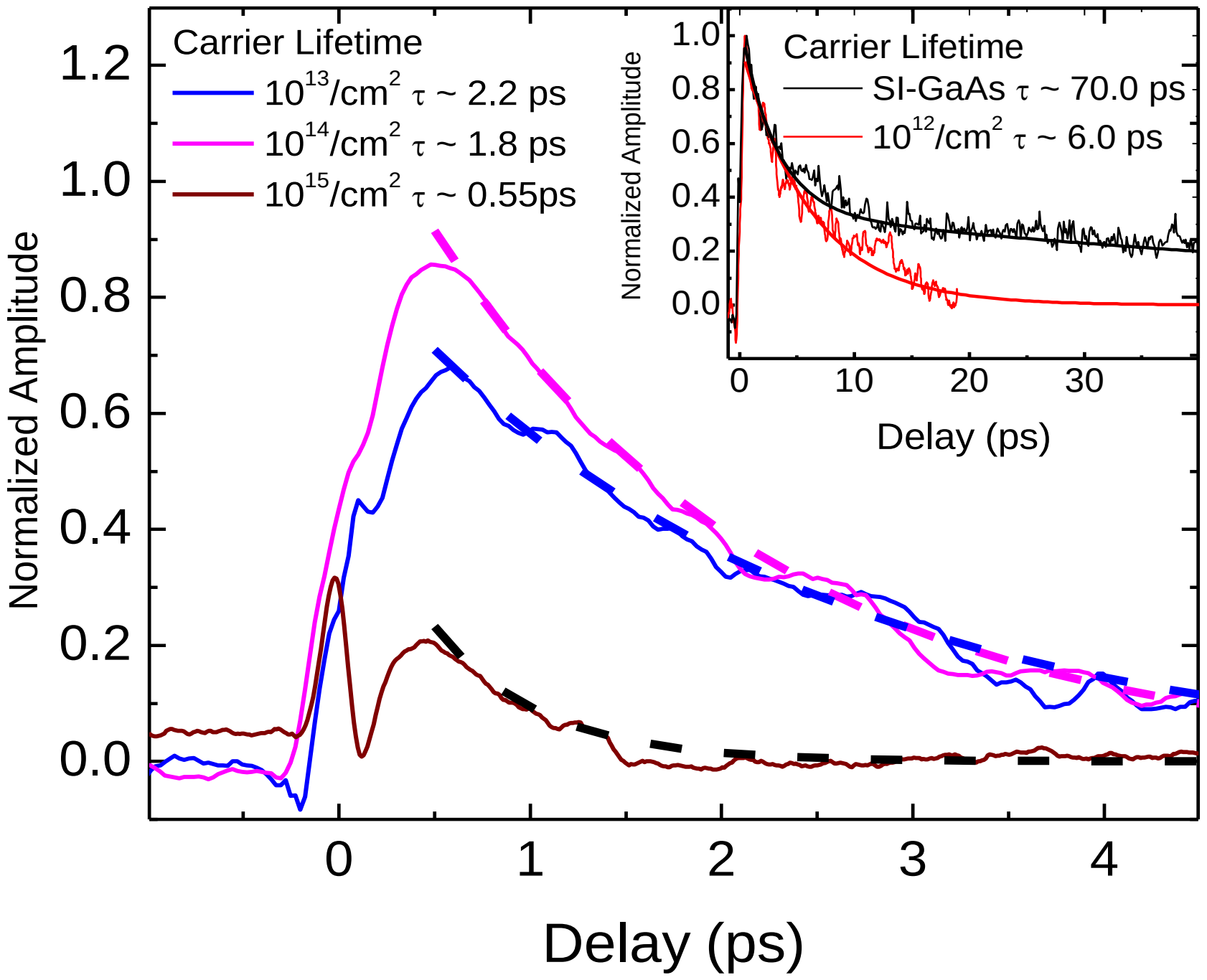

Figure3. Optical IR time resolved pump-probe reflection curves of non-irradiated and irradiated SI-GaAs. The dash-lines are fits to single exponential decay τ. The inset shows non-irradiated sample with a slow fall (~70 ps) in the signal while the irradiated sample shows comparatively rapid fall (~6 ps).

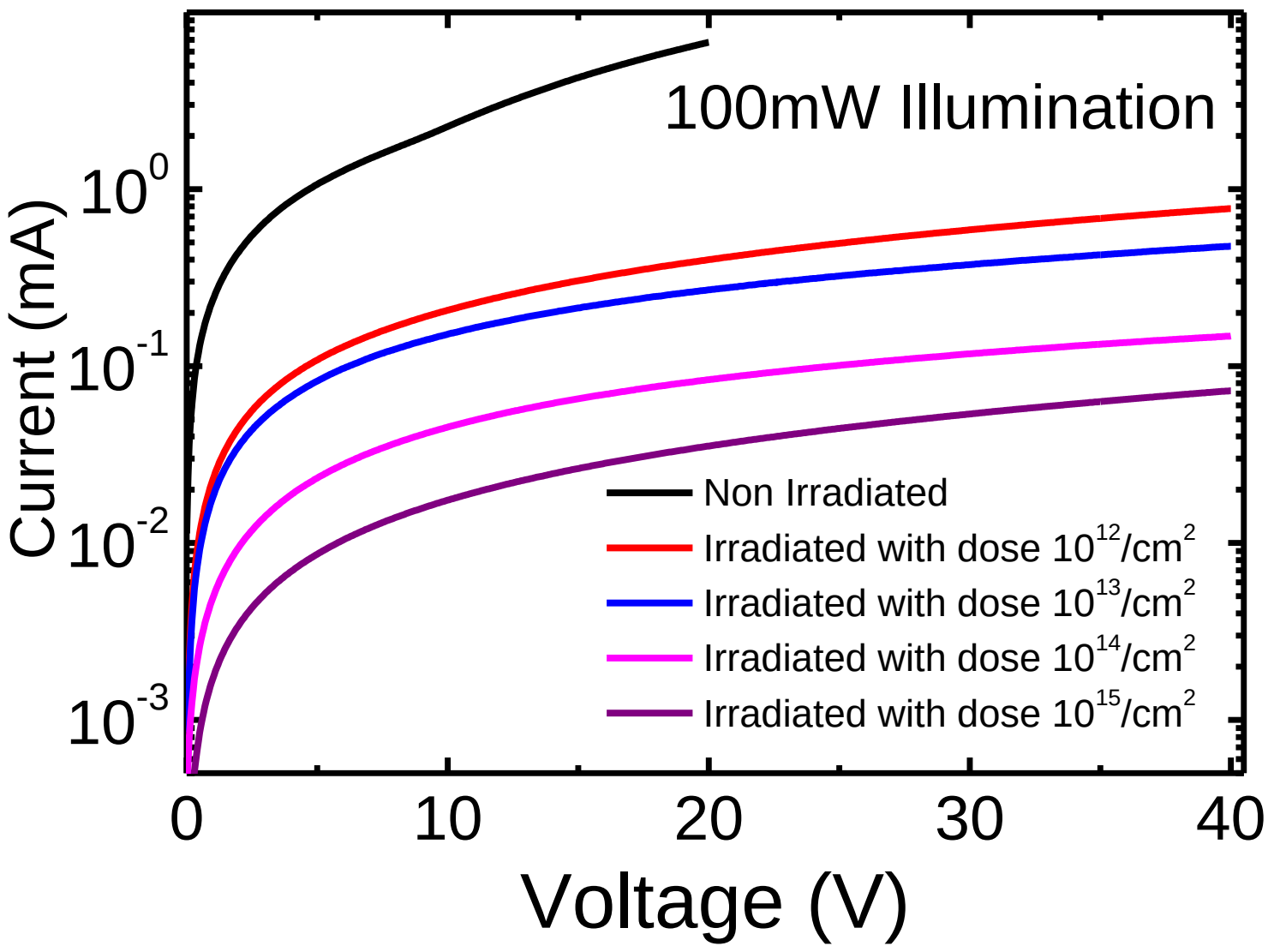

Figure4. Current-Voltage characteristics of PCEs fabricated on irradiated and non-irradiated SI-GaAs substrate under 100 mW IR-pulsed laser illumination for various C-ion irradiation doses. The current for highest irradiation dose sample is almost 2 orders of magnitude smaller than the non-irradiated sample.

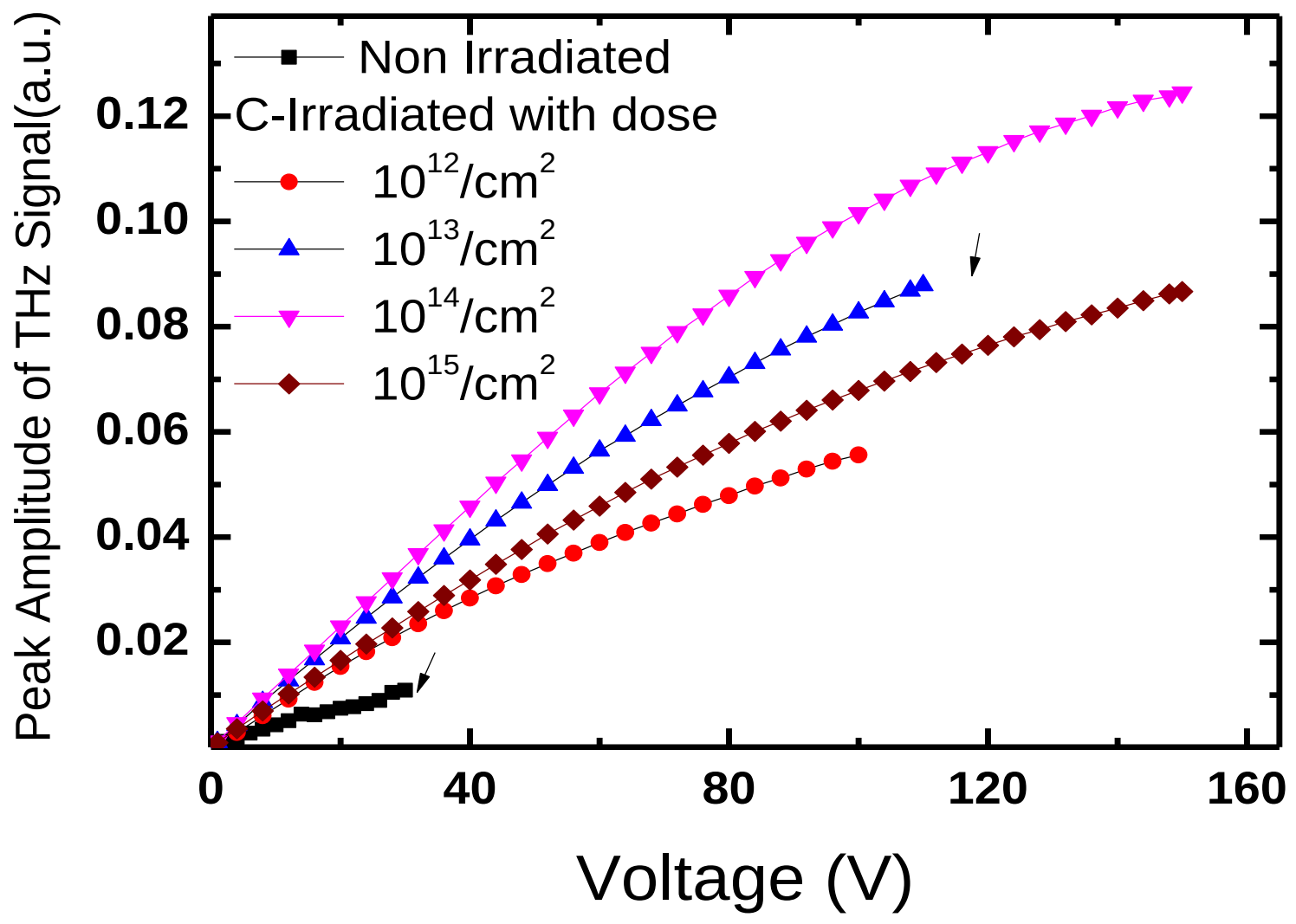

Figure5. Variation of emitted THz pulse amplitude with applied bias from non-irradiated and Carbon-irradiated SI-GaAs sources. The breakdown voltage for non-irradiated sample is the lowest ~30 V and for the higher dosage is more than 150 V.